\documentclass[final,1p,times,authoryear]{elsarticle}

\usepackage[utf8]{inputenc} 
\usepackage[T1]{fontenc}    
\usepackage{hyperref}       
\usepackage{url}            
\usepackage{booktabs}       
\usepackage{amsfonts}       
\usepackage{nicefrac}       
\usepackage{microtype}      
\usepackage{xcolor}         
\usepackage{multirow}  

\usepackage{amsmath}   
\usepackage{amssymb}   
\usepackage{amsthm}    

\newtheorem{theorem}{Theorem}[section]  
\newtheorem{definition}{Definition}

\usepackage{cite}
\usepackage{amsmath,amssymb,amsfonts}
\usepackage{algorithmic}
\usepackage{graphicx}
\usepackage{epstopdf}
\usepackage{textcomp}
\usepackage{xcolor}
\usepackage{adjustbox}
\usepackage{color}
\usepackage{booktabs}
\usepackage{tabularx}  
\usepackage{caption}
\usepackage{threeparttable}  
\usepackage{float}  
\usepackage{array}  
\usepackage{multirow}  
\usepackage{color}
\usepackage{amsthm}
\usepackage{afterpage}

\usepackage{graphicx} 
\usepackage{amssymb}
\usepackage{amsthm}

\usepackage[utf8]{inputenc} 
\usepackage[T1]{fontenc}    
\usepackage{physics} 
\usepackage{amssymb} 
\usepackage{hyperref}       
\usepackage{enumitem} 
\usepackage{url}            
\usepackage{booktabs}       
\usepackage{amsfonts}       
\usepackage{nicefrac}       
\usepackage{microtype}      
\usepackage{xcolor}         
\usepackage{amsmath} 
\usepackage{xcolor}
\usepackage{wrapfig}

\usepackage{longtable}
\usepackage{booktabs}

\usepackage{booktabs} 
\usepackage{multirow} 

\setlist{nosep}

\usepackage{xcolor}
\usepackage[dvipsnames]{xcolor}

\usepackage{xcolor}

\usepackage{amssymb}
\usepackage{amsmath}

\journal{Clinical Neurophysiology}

\begin{document}

\begin{frontmatter}



\title{Complexity and Stability of Neural Activity Across Aging and Neurodegenerative Disease
} 

\author{Junjie Yu$^{a,b,*}$ \quad Jianyu Zhang$^{a,*}$ 
Zian Pei$^{b}$ \quad Xue Shi$^{b}$ \\
\quad Yumei Liu$^{b}$ 
\quad Xin Jiang$^{c}$
\quad Quanying Liu$^{a}$
\quad Yi Guo$^{b, d, e, f, \dag}$} 




\affiliation{organization={Department of Biomedical Engineering, Southern University of Science and Technology}, city={Shenzhen}, country={China}}

\affiliation{organization={Department of Neurology, Shenzhen People's Hospital, The Second Affiliated Hospital of Jinan University, The First Affiliated Hospital of Southern University of Science and Technology}, city={Shenzhen}, country={China}}

\affiliation{organization={Department of Geriatric, Shenzhen People's Hospital, The Second Affiliated Hospital of Jinan University, The First Affiliated Hospital of Southern University of Science and Technology}, city={Shenzhen}, country={China}}

\affiliation{organization={School of Medicine, Southern University of Science and Technology}, city={Shenzhen}, country={China}}

\affiliation{organization={Department of Neurology, Tianjin Huanhu Hospital}, city={Tianjin}, country={China}}

\affiliation{organization={Institute of Neurological and Psychiatric Disorders, Shenzhen Bay Laboratory}, city={Shenzhen}, country={China}}

\footnotetext[1]{$^{*}$These authors contributed equally to this work.}
\footnotetext[2]{$^{\dag}$Corresponding author.}

\begin{abstract}
\textbf{Objective}: EEG signals fluctuate continuously even within a fixed cognitive state, but an important question is whether the brain still reuses similar activity patterns to represent information over time. \textbf{Methods}: To address this, we model EEG as distributions of windowed activity patterns and quantify their temporal stability using Wasserstein distance, while intrinsic dimensionality captures representational complexity. \textbf{Results}: Across multi-task, lifespan, and clinical EEG datasets, we find that neural representations show constrained, condition-specific stability rather than unconstrained drift. Higher intrinsic dimensionality is consistently associated with lower stability, suggesting that richer representational spaces are less reproducible over time. Both measures exhibit reproducible spatial organization, with posterior regions showing higher dimensionality and lower stability than frontal regions. Healthy aging is characterized by increased dimensionality and reduced stability, whereas mild cognitive impairment and Alzheimer’s disease show a joint collapse of both. \textbf{Conclusions}: These findings provide a distribution-level framework for understanding neural stability across cognition, aging, and disease. \textbf{Significance}: This framework offers a principled approach to quantifying neural representational stability, with potential utility as a sensitive biomarker for tracking cognitive aging and neurodegeneration in clinical settings.
\end{abstract}



\begin{keyword}



EEG, Aging, Neurodegenerative Diseases, Alzheimer's Disease

\end{keyword}

\end{frontmatter}

\section{Introduction}





Brain activity is inherently dynamic. Even during quiet rest or repeated performance of the same task, EEG signals continue to fluctuate over time \citep{driscoll2022representational, adam2022dynamics, druckmann2012neuronal, orban2016neural}. The key question, however, is not simply whether neural signals vary, but whether the brain continues to recruit similar activity patterns when the cognitive state remains unchanged. This distinction is important because the stable reuse of activity patterns may reflect reliable and efficient neural coding, whereas reduced stability may indicate changes in how the brain organizes and represents information. Understanding this balance is particularly relevant for characterizing brain function across healthy aging and neurodegenerative disease \citep{dinstein2015neural, renart2014variability, raviv2022variability, grady2014understanding, anderson2016cognitive, burton2006intraindividual}.

\begin{figure}[htbp]
  \centering
  \includegraphics[width=0.99\linewidth]{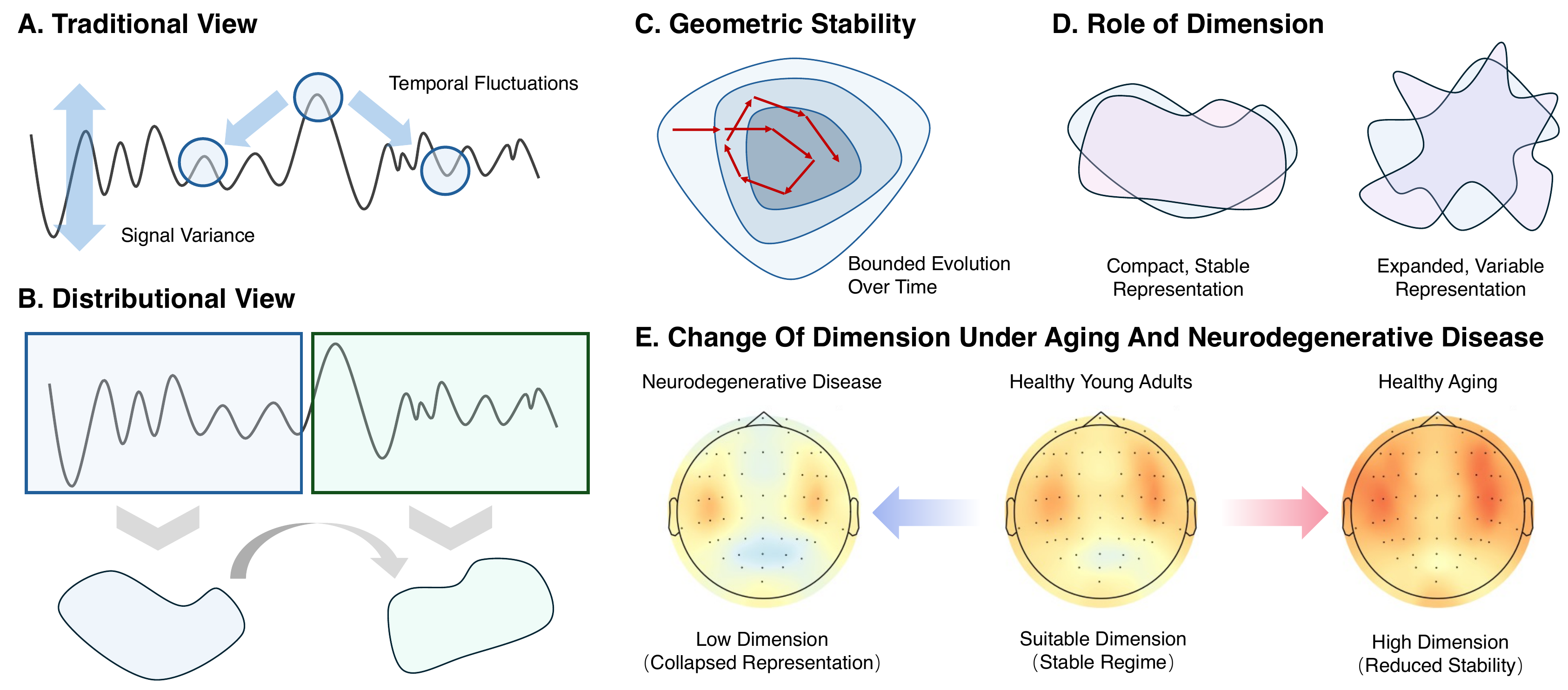}
  \caption{\textbf{Geometric framework for analyzing neural population stability.} \textbf{(A)} Traditional Signal View: Moment-to-moment signal variability makes it difficult to distinguish stable states from noise. \textbf{(B)} Distributional View: Neural activity is conceptualized as probability distributions (clouds) in a high-dimensional space. \textbf{(C)} Geometric Stability: Stability is defined by constrained distributional evolution, allowing for signal dynamics without representation collapse. \textbf{(D)} Role of Intrinsic Dimensionality: Intrinsic dimensionality constrains stability; low-dimensional manifolds (blue/purple) limit displacement, while high dimensions (orange/red) increase instability. \textbf{(E)} Unified Lifespan and Disease Framework: Healthy function occupies a mid-dimensional stable regime. Aging shifts toward high-dimensional instability, while neurodegenerative disease leads to low-dimensional representational collapse.}
  \label{fig:motivation}
\end{figure}

Most EEG studies have described temporal changes using signal-level measures such as variance \citep{clopath2017variance}, spectral power \citep{knyazev2009cortical}, entropy \citep{ignaccolo2010dynamics, lau2022brain}, or temporal correlation \citep{benayoun2010eeg, berthouze2010human} (Figure \ref{fig:motivation}A). These measures are useful and have revealed important changes in aging and disease. However, they mainly describe how much the signal fluctuates, not whether the same underlying activity pattern is being used again and again. Two EEG segments may differ in amplitude or complexity while still reflecting a similar functional state, whereas similar signal statistics do not guarantee that the brain is using the same pattern of activity. For clinical interpretation, this difference is important: a brain can be variable without being disorganized, and it can appear stable because it is pathologically restricted rather than healthy.

To address this gap, we treat brain activity as a collection of multichannel patterns sampled over time from an underlying patterns of possible states (Figure \ref{fig:motivation}B). In this view, stability means that patterns observed in one time segment resemble those observed in another segment under the same condition. We quantify this similarity using Wasserstein distance, which measures how different two sets of activity patterns are. This allows us to move beyond moment-to-moment signal fluctuation and ask a more clinically and biologically meaningful question: does the brain keep returning to a similar set of activity states, or does that set shift over time?

A central idea of this study is that the answer may depend on how broad the range of EEG activity patterns is. If brain activity is confined to a relatively limited set of patterns, then similar patterns are more likely to reappear over time. If activity can vary along many meaningful directions, the observed patterns may become less reproducible. We describe this property using intrinsic dimensionality (Figure \ref{fig:motivation}C--D), which can be understood intuitively as the number of meaningful ways brain activity can vary. In this sense, intrinsic dimensionality provides a simple geometric explanation for why some brain states are more temporally stable than others.

We tested this framework in large-scale human EEG datasets spanning cognitive tasks, healthy aging, mild cognitive impairment, and Alzheimer's disease. These settings address complementary aspects of brain organization. Cognitive tasks probe how neural patterns adapt across mental states. Healthy aging captures gradual changes in the balance between flexibility and stability across the adult lifespan. Mild cognitive impairment and Alzheimer's disease provide a clinically important contrast, allowing us to ask whether pathological brain activity is simply more variable, or whether it becomes abnormally restricted to a narrower range of usable patterns.

Across these analyses, we found that neural activity under fixed conditions does not drift in an unconstrained way. Instead, it shows bounded and condition-specific stability over time. We further found that this stability is closely linked to intrinsic dimensionality across brain regions and conditions. Healthy aging was associated with higher dimensionality and lower stability, suggesting a broader but less efficient use of available neural patterns. By contrast, mild cognitive impairment and Alzheimer's disease showed lower dimensionality together with abnormally increased stability, consistent with a pathological narrowing of the brain's representational pattern (Figure \ref{fig:motivation}E).

This study makes three main contributions:
\begin{enumerate}
    \item We provide a distribution-based framework for measuring whether brain activity patterns remain similar over time under fixed conditions.
    \item We show that intrinsic dimensionality, which captures the richness of brain activity patterns, helps explain their stability across brain regions and conditions.
    \item We demonstrate that aging and neurodegenerative disease reshape brain activity in opposite directions, with aging associated with higher dimensionality and lower stability, and disease associated with lower dimensionality and abnormally elevated stability.
\end{enumerate}

Together, these findings provide a clinically interpretable framework for understanding how brain activity is organized over time and how this organization changes in aging and neurodegenerative disease.

\section{Preliminaries and Technical Background}

\subsection{EEG Datasets and Participants}

The present study analyzes three complementary EEG datasets, each chosen to examine neural representational stability across different cognitive conditions, temporal scales, and population groups.

\paragraph{\textbf{Multi-task EEG dataset}} This publicly available dataset \citep{wang2022test} includes 60 healthy university students (ages 18--28, all right-handed, no history of neurological or psychiatric disorders). The dataset includes two resting-state recordings (eyes open and eyes closed, 5 minutes each) and three internally driven cognitive tasks under eyes-closed conditions: episodic memory recall, mental singing, and continuous mental arithmetic. In the present analysis, we primarily used the eyes-closed resting-state and task recordings to examine differences in EEG activity patterns across cognitive conditions while reducing variability related to eye state.

\paragraph{\textbf{Lifespan EEG dataset}} This dataset comprises resting-state EEG recordings from healthy participants across a wide age range, collected at Shenzhen People's Hospital. It serves as a normative baseline to characterize the structure and complexity of neural representations across the adult lifespan. The total sample size includes 365 participants, with balanced representation across age groups.

\paragraph{\textbf{Clinical EEG dataset}} This dataset contains 374 participants: 136 healthy controls (HC), 129 individuals with mild cognitive impairment (MCI), and 109 patients with clinically diagnosed Alzheimer's disease (AD). All participants were assessed and grouped according to the 2011 NIA-AA criteria \citep{jack2024revised}, beginning with MoCA screening (threshold score 26) followed by clinical evaluation by neurologists to distinguish MCI and AD. EEG recordings were collected in a standardized resting-state protocol across all groups and preprocessing pipeline, including re-referencing, artifact removal via ICA, and manual inspection, minimizing system-level biases across groups. This dataset allows examination of how neurodegenerative processes alter the geometric structure and stability of neural representations.

Together, these three datasets provide complementary views on neural population activity: the multi-task dataset captures repeated measures within individuals under controlled cognitive states, the lifespan dataset characterizes normative variability across ages, and the clinical dataset enables assessment of pathological alterations. The continuous EEG recordings, combined with relatively weak task constraints, make these datasets suitable for a distributional analysis of neural activity as repeated samples from underlying representational spaces.

\subsubsection{EEG as a collection of windowed samples}

To obtain a common representation across datasets, all EEG recordings were analyzed using a harmonized preprocessing framework. Signals underwent ICA-based artifact removal and were downsampled to 100 Hz before further analysis. Our analysis was performed separately for each EEG channel, allowing channel-wise characterization of temporal pattern stability across subjects and conditions.

For a given subject, condition, and channel, the continuous EEG time series was divided into non-overlapping 1-second windows. At a sampling rate of 100 Hz, each window was represented as a 100-dimensional vector of signal values after standardization. Thus, each channel produced a sequence of short temporal patterns sampled across time.

We treated the set of windows from a given channel and condition as an empirical sample from an underlying distribution of neural activity patterns. A 1-second window provides a simple and interpretable unit that captures short-term temporal structure while preserving sufficient samples for distributional analysis. This representation shifts the analysis from isolated time points to the distribution of local activity patterns, allowing us to ask whether the same channel repeatedly expresses similar patterns over time under a fixed cognitive or clinical state.

\subsubsection{Intrinsic dimensionality}

To characterize the complexity of the windowed EEG patterns, we estimated intrinsic dimensionality. Importantly, intrinsic dimensionality does not refer to the raw dimensionality of the observation vector itself (here, 100 samples per window), but to the effective number of degrees of freedom needed to describe how the observed patterns vary. A lower value indicates that the samples lie near a relatively compact and constrained structure, whereas a higher value indicates that they spread across a broader and more complex region of the space.

In the present study, intrinsic dimensionality serves as a summary of how richly EEG activity varies within a channel and condition. Intuitively, it reflects how many meaningful ways the short-term activity pattern can change over time. We estimated intrinsic dimensionality using the maximum likelihood estimator of \citet{levina2004maximum}, implemented in the \texttt{scikit-dimension} (\texttt{skdim}) package \citep{bac2021scikit}, which provides a standard neighborhood-based approach for recovering local low-dimensional structure from high-dimensional samples. Unless otherwise noted, all analyses used a neighborhood size of $K=50$. To evaluate the sensitivity of the findings to this parameter, supplementary analyses using $K=10$ and $K=100$ are provided in the \ref{appendix:multitask_validK} -- \ref{appendix:disease_validK}.

\subsubsection{Wasserstein distance}

To compare EEG activity patterns across time segments, we used Wasserstein distance. Given two empirical distributions, Wasserstein distance measures the minimal transport cost required to transform one distribution into the other. Smaller values indicate that the two sets of samples are more similar, whereas larger values indicate greater distributional dissimilarity.

In our framework, Wasserstein distance provides a direct measure of temporal stability within a fixed condition. If two time segments from the same channel express similar patterns, their empirical distributions should have a small Wasserstein distance. If the underlying activity patterns shift over time, the distance should increase. Unlike pointwise signal differences or summary statistics based only on amplitude or variance, Wasserstein distance compares the geometry of the full empirical distributions. We computed these distances using the \texttt{POT} (\texttt{Python Optimal Transport}) package \citep{flamary2021pot} and used them as our measure of distributional stability.

\subsubsection{Link between dimensionality and stability}

Our analysis is based on the idea that the geometric complexity of EEG activity constrains how stable its empirical distribution appears over time. When windowed samples occupy a relatively compact low-dimensional structure, repeated temporal segments are more likely to yield similar empirical distributions \citep{weed2019sharp}. By contrast, when samples spread across a broader high-dimensional space, finite observations from different time segments are more likely to show greater distributional variation.

This link should be understood as a statistical tendency rather than a deterministic rule: intrinsic dimensionality does not by itself determine stability, but it provides a principled constraint on how reproducible short-term activity patterns can be under repeated sampling. This motivates our joint analysis of intrinsic dimensionality and Wasserstein distance as complementary descriptors of neural organization. Additional theoretical background is provided in \ref{app:theory}.

\section{Experiments and Results}

Using the geometric framework described above, we examined three related questions:
(i) whether neural representations remain stable over time under fixed cognitive conditions,
(ii) whether intrinsic dimensionality is associated with representational stability, and
(iii) how these two properties change with healthy aging and neurodegenerative disease.

\subsection{Neural representations exhibit constrained distributional evolution within fixed cognitive conditions}

Before analyzing distributional stability across brain regions and cognitive conditions, we first asked whether within-condition stability can be meaningfully characterized. If neural activity under a fixed cognitive condition were to drift continuously over time, then any estimate of stability would mainly reflect elapsed time rather than a stable property of the underlying representation. We therefore tested whether neural activity within a fixed cognitive condition shows progressive distributional drift or instead remains within a condition-specific distributional regime.

To test this, we divided each recording into successive non-overlapping time segments and computed the Wasserstein distance between each segment and an initial reference segment from the same condition.

\begin{figure}[!t]
  \centering
  \includegraphics[width=0.99\linewidth]{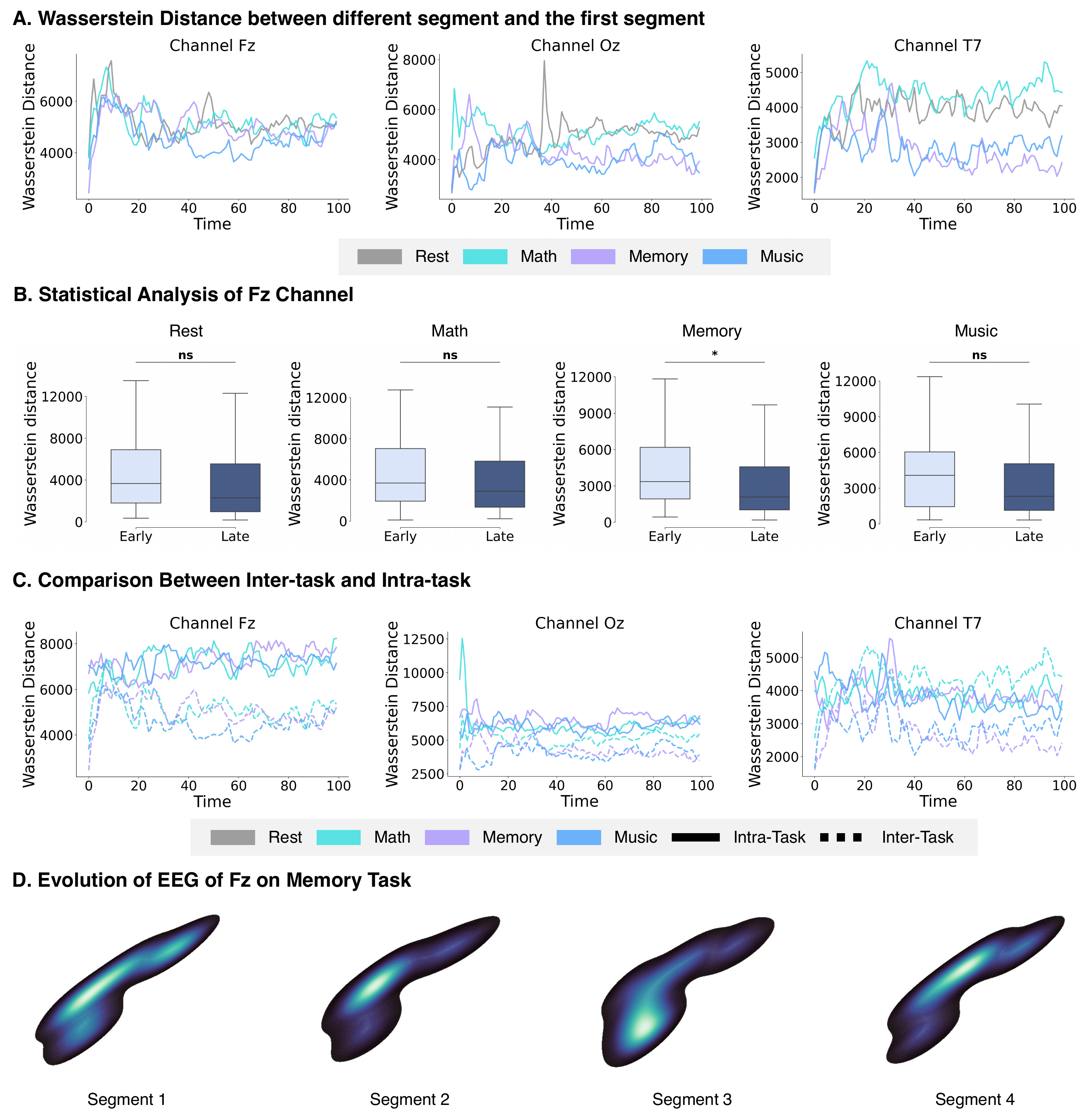}
  \caption{\textbf{Constrained distributional evolution of neural representations within fixed cognitive conditions.}
  (A) Wasserstein distance between empirical distributions from successive time segments and the initial reference segment shows bounded, non-monotonic fluctuations over time.
  (B) Group-level comparison of early and late segments relative to the same initial distribution shows no significant difference in Wasserstein distance.
  (C) Within-condition Wasserstein distances are consistently smaller than cross-condition distances, indicating stronger distributional similarity within the same cognitive condition.
  (D) Visualization of empirical distributions for the Fz channel during the scene memory task shows that samples from different time segments remain clustered within a compact region of the reduced representational space.}
  \label{fig:wassersteinDynamics}
\end{figure}

Across subjects and task conditions, Wasserstein distance showed non-monotonic fluctuations but remained bounded over time, with no evidence of cumulative growth (Fig.~\ref{fig:wassersteinDynamics}A). Representational displacement therefore neither increased progressively, as expected under systematic drift, nor converged toward zero, as expected if neural activity collapsed onto a fixed configuration.

To further test this observation, we compared early and late segments relative to the same initial reference distribution. Early and late Wasserstein distances did not differ significantly across subjects (Fig.~\ref{fig:wassersteinDynamics}B), providing additional evidence against progressive representational drift.

We next asked whether these constrained fluctuations were condition-specific or simply reflected generic temporal dependence. To test this, we compared within-condition Wasserstein distances with cross-condition distances obtained by pairing the reference segment with segments from other cognitive conditions. Within-condition distances were consistently smaller than cross-condition distances over time (Fig.~\ref{fig:wassersteinDynamics}C). This indicates that neural activity remains closer to its own condition-specific distribution than to distributions associated with other conditions.

As an illustrative example, we visualized the empirical distributions for a representative frontal channel (Fz) during the scene memory task. Although the signal varied substantially over time, distributions from different segments remained clustered within a compact region of the reduced representational space (Fig.~\ref{fig:wassersteinDynamics}D).

Together, these findings argue against unconstrained drift, collapse to a fixed point, and cross-condition mixing as dominant explanations of neural dynamics under fixed cognitive conditions. Instead, neural activity remains within condition-specific regions of representational space without evidence of progressive drift. Notably, the scale of these within-condition fluctuations varied systematically across cortical regions, motivating the next analysis of their geometric determinants. More results are provided in \ref{appendix:constrained_channel}.

\subsection{Intrinsic dimensionality and distributional stability exhibit stable spatial organization across cognitive conditions}

Having shown that within-condition Wasserstein distance captures meaningful, non-cumulative variation over time, we next asked how this quantity is organized across the cortex. Specifically, we examined whether intrinsic dimensionality and within-condition Wasserstein distance exhibit reproducible whole-brain topographies across resting-state and task conditions.

\begin{figure}[htbp]
  \centering
  \includegraphics[width=0.97\linewidth]{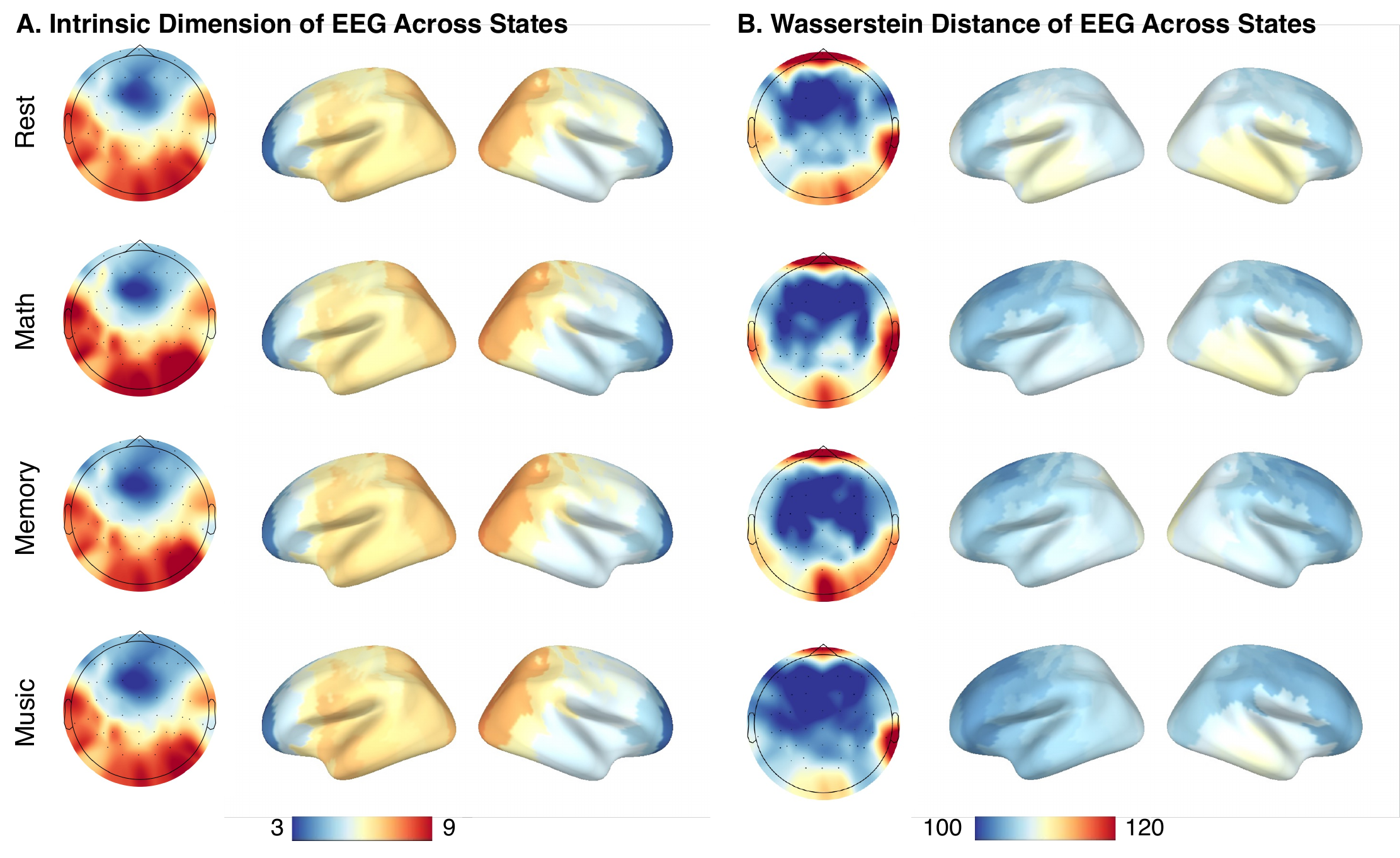}
    \caption{\textbf{Intrinsic dimensionality and distributional stability exhibit reproducible spatial organization across cognitive conditions.}
    (A) Whole-brain topography of intrinsic dimensionality across resting-state and task conditions, showing a posterior--anterior gradient with higher values in posterior regions and lower values in frontal regions.
    (B) Whole-brain topography of within-condition Wasserstein distance across the same conditions, showing a similar spatial gradient, with larger within-condition displacement in posterior regions and smaller displacement in frontal regions.}
  \label{fig:correlationMultiTask}
\end{figure}

To address this question, we estimated intrinsic dimensionality and within-condition Wasserstein distance for each EEG channel in the resting-state condition and in all task conditions. Across conditions, intrinsic dimensionality showed a robust posterior--anterior gradient (Fig.~\ref{fig:correlationMultiTask}A), with occipital and temporal regions showing higher values and frontal regions showing markedly lower values. This pattern was consistently observed across tasks, indicating that intrinsic dimensionality follows a stable large-scale cortical organization.

A similar spatial pattern was observed for within-condition Wasserstein distance (Fig.~\ref{fig:correlationMultiTask}B). Posterior channels generally exhibited larger representational displacement between time segments from the same condition, corresponding to lower distributional stability, whereas frontal regions showed smaller distances and greater overlap between empirical distributions. This pattern remained broadly similar across cognitive conditions, indicating that distributional stability is likewise organized along a stable cortical gradient.

Together, these findings show that both intrinsic dimensionality and within-condition Wasserstein distance exhibit reproducible whole-brain organization across resting-state and task conditions. Rather than being dominated by condition-specific fluctuations, both measures follow characteristic spatial gradients that are preserved across conditions. The similarity of these large-scale patterns further suggests that intrinsic dimensionality and distributional stability are systematically related across cortical regions, which we examine next.

\subsection{Healthy aging is associated with representational expansion and reduced distributional stability}

Having shown that intrinsic dimensionality and distributional stability exhibit reproducible spatial organization across cognitive conditions, we next asked how these geometric properties change across the adult lifespan. If aging systematically reshapes this geometric framework, then intrinsic dimensionality and within-condition Wasserstein distance should show parallel changes with age.

\begin{figure}[!t]
  \centering
  \includegraphics[width=0.95\linewidth]{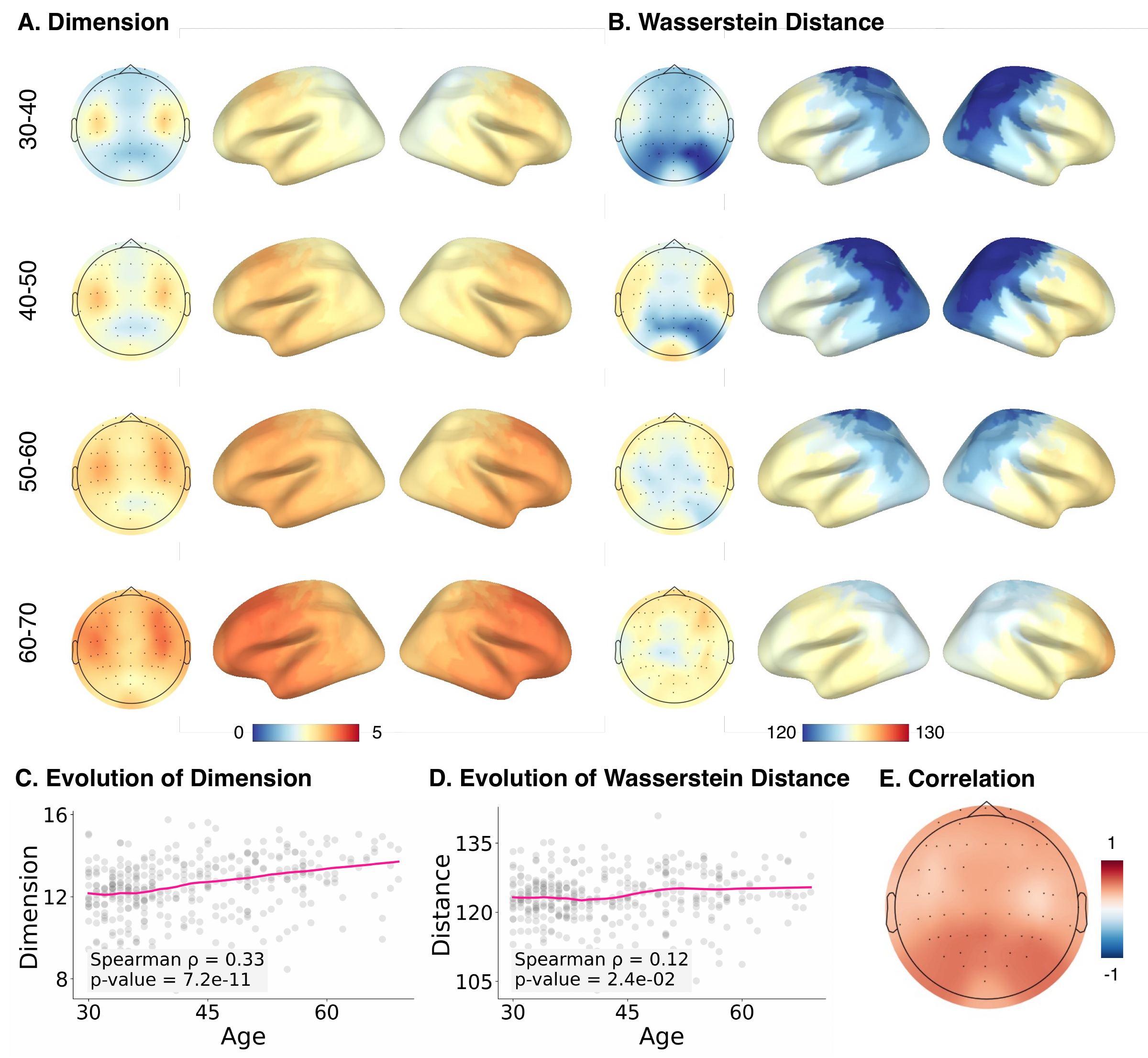}
    \caption{\textbf{Healthy aging is associated with representational expansion and reduced distributional stability.}
    (A) Whole-brain topography of intrinsic dimensionality across age groups.
    (B) Whole-brain topography of within-condition Wasserstein distance across age groups.
    (C) Intrinsic dimensionality increases significantly with age.
    (D) Within-condition Wasserstein distance increases significantly with age.
    (E) Channel-wise association between intrinsic dimensionality and within-condition Wasserstein distance across cortical regions, showing significant positive correlations after FDR correction.}
  \label{fig:aging}
\end{figure}

Whole-brain topographies across age groups showed widespread age-related increases in intrinsic dimensionality (Fig.~\ref{fig:aging}A) and within-condition Wasserstein distance (Fig.~\ref{fig:aging}B). These effects were spatially broad and largely preserved the posterior--anterior gradient observed in younger adults, suggesting a global shift in representational geometry rather than region-specific reorganization.

Quantitatively, both intrinsic dimensionality and within-condition Wasserstein distance increased significantly with age (Fig.~\ref{fig:aging}C--D). Accordingly, neural activity in older adults showed higher intrinsic dimensionality and larger displacement between time segments from the same resting-state condition, indicating reduced distributional stability over time.

Critically, intrinsic dimensionality and within-condition Wasserstein distance were significantly positively correlated across channels, and this association survived FDR correction (Fig.~\ref{fig:aging}E). This indicates that cortical regions with higher intrinsic dimensionality also show reduced distributional stability during healthy aging.

Together, these findings suggest that healthy aging is characterized by representational expansion. As neural activity occupies a larger effective volume of representational space, it reflects a broader and more differentiated set of activity patterns. This is associated with greater divergence between empirical distributions sampled over time and reduced overlap among neural states expressed within the same condition. In this sense, healthy aging is associated not with representational collapse, but with a shift toward a higher-dimensional and less compact representational regime.

\subsection{Neurodegenerative disease is associated with collapse and rigidification of representational geometry}

Having shown that healthy aging is associated with representational expansion, we next examined how intrinsic dimensionality and distributional stability jointly change in neurodegenerative disease. Specifically, we asked whether these two geometric properties are systematically altered in patient groups, and whether their association is preserved under pathological conditions.

\begin{figure}[!t]
  \centering
  \includegraphics[width=0.98\linewidth]{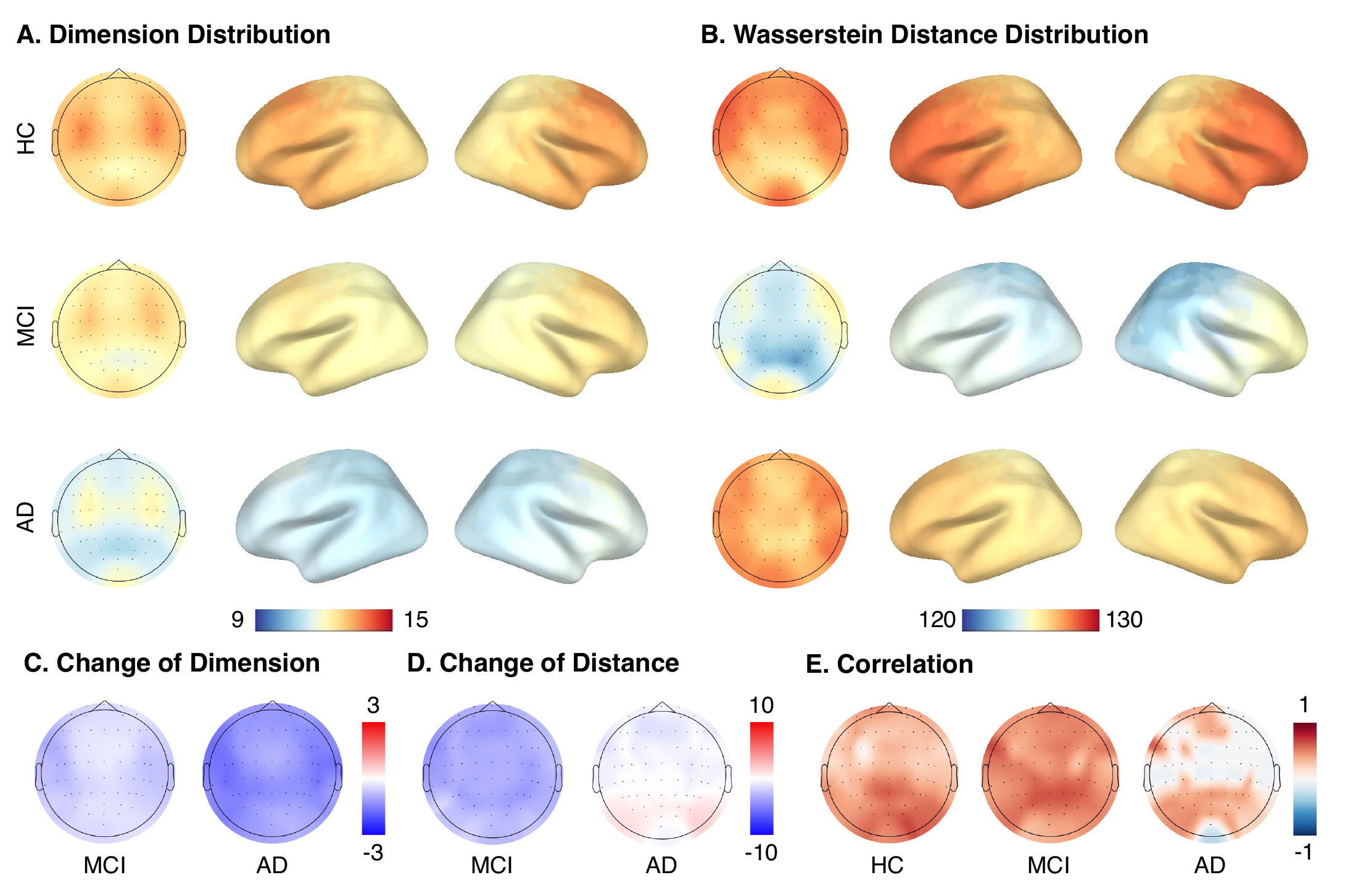}
    \caption{\textbf{Neurodegenerative disease is associated with collapse and rigidification of representational geometry.}
    (A) Whole-brain topography of intrinsic dimensionality in healthy controls (HC), mild cognitive impairment (MCI), and Alzheimer's disease (AD).
    (B) Whole-brain topography of within-condition Wasserstein distance across groups.
    (C) Relative to HC, both MCI and AD show spatially widespread reductions in intrinsic dimensionality.
    (D) Relative to HC, within-condition Wasserstein distance is also significantly reduced across broad cortical regions, although reductions in AD are less pronounced in occipital areas.
    (E) Channel-wise association between intrinsic dimensionality and within-condition Wasserstein distance across cortical regions within each group. Significant positive associations are widespread in HC and MCI, whereas AD shows a reduced spatial extent of significant correlations after FDR correction.}
  \label{fig:disease}
\end{figure}

Using resting-state EEG recordings, we compared healthy controls (HC) with individuals diagnosed with mild cognitive impairment (MCI) and Alzheimer's disease (AD). Whole-brain topographies showed that both clinical groups exhibited lower intrinsic dimensionality than healthy controls (Fig.~\ref{fig:disease}A). In parallel, within-condition Wasserstein distance was also reduced in MCI and AD relative to HC (Fig.~\ref{fig:disease}B), indicating smaller distributional displacement between time segments from the same resting-state condition.

We next quantified these group differences directly. Relative to HC, both MCI and AD exhibited spatially widespread reductions in intrinsic dimensionality across the cortex (Fig.~\ref{fig:disease}C), indicating large-scale collapse of representational geometry. Within-condition Wasserstein distance was likewise significantly reduced across broad cortical regions in both clinical groups (Fig.~\ref{fig:disease}D), although in AD the reduction was less pronounced in occipital regions. Thus, neurodegenerative disease is associated not only with lower-dimensional neural representations, but also with a more rigid and constrained representational regime.

We then asked whether the association between intrinsic dimensionality and within-condition Wasserstein distance was preserved across groups. In HC and MCI, these two measures were significantly positively associated across a broad set of cortical regions after FDR correction (Fig.~\ref{fig:disease}E). In AD, a positive association was still present, but the spatial extent of significant correlations was substantially reduced. These results suggest that neurodegenerative disease does not fully abolish the geometric relationship between dimensionality and distributional stability, but is associated with increasing compression of neural activity into a low-dimensional and over-stable representational regime.

Importantly, reduced within-condition Wasserstein distance in neurodegenerative disease should not be interpreted as preserved functional stability. Rather, in the context of collapsed intrinsic dimensionality, it reflects pathological over-stability: neural population activity becomes confined to a narrow representational subspace and repeatedly expresses highly similar configurations over time. Reduced displacement therefore indicates a loss of representational flexibility and a reduced capacity to express distinct encoding patterns, rather than improved maintenance of cognitive representations.

Taken together with the aging results, these findings indicate that healthy aging and neurodegeneration correspond to opposite shifts along a shared geometric axis. In healthy aging, neural representations expand into a higher-dimensional and less compact regime. In contrast, neurodegenerative disease is characterized by collapse and rigidification of representational geometry, accompanied by a reduced range of viable activity patterns and a diminished capacity to express informative or differentiated neural states. Across both regimes, intrinsic dimensionality remains a key geometric constraint on distributional stability, providing a unified account that distinguishes representational expansion from pathological representational collapse.

\section{Discussion and Conclusion}

In this study, we proposed a distributional and geometric framework for characterizing the temporal stability of neural population activity. By modeling neural activity as empirical distributions evolving in representational space, we move beyond signal-level variability and quantify stability at the level of population representations. This perspective complements conventional measures such as variance, spectral power, entropy, and temporal correlation, which are informative about signal fluctuation but do not directly capture the reuse of population-level representational structure over time \citep{knyazev2009cortical, berthouze2010human}. More broadly, it aligns with growing interest in representational geometry and low-dimensional population organization as fundamental principles of neural computation \citep{kriegeskorte2013representational, gao2015simplicity, saxena2019towards}.

Our results show that neural representations are neither static nor unconstrained. Rather than drifting arbitrarily or collapsing to fixed configurations, neural population activity repeatedly samples from structured regions of representational space under fixed cognitive conditions, generating constrained but non-identical empirical distributions over time \citep{renart2014variability, orban2016neural}. In this sense, neural stability is better understood not as immobility, but as constrained distributional evolution within a structured geometric substrate.

A central finding of this work is that intrinsic dimensionality and distributional stability are closely linked across cortical regions, cognitive conditions, healthy aging, and neurodegenerative disease. Regions operating in higher-dimensional regimes consistently exhibit larger within-condition Wasserstein distances, whereas lower-dimensional regions show more compact and overlapping representational distributions. Importantly, this relationship should not be interpreted as a trivial reduction of neural dynamics in low-dimensional regions. Instead, intrinsic dimensionality appears to impose a geometric constraint on the range of accessible neural configurations, thereby shaping how far empirical distributions can diverge over time. From this perspective, dimensionality can be viewed as a key geometric variable governing representational exploration and stability.

This geometric framework also provides a unified account of lifespan- and disease-related changes in neural representation. In healthy aging, intrinsic dimensionality and within-condition representational displacement increase together, indicating that neural activity occupies a larger effective volume of representational space. These patterns are broadly consistent with previous work showing that healthy aging is associated with altered neural variability and reduced representational specificity, whereas neurodegenerative disease is associated with reduced neural complexity and diminished dynamical flexibility \citep{zappasodi2015age, besthorn1997discrimination, smits2016electroencephalographic, namazi2018age, yagyu1997global, yang2019m}. Importantly, this reduction in Wasserstein distance should not be interpreted as preserved functional stability, since healthy neural function depends not only on persistence but also on sufficient variability and dynamical flexibility \citep{garrett2011importance, deco2011dynamical, tognoli2014metastable}. Rather, it likely reflects pathological over-stability, consistent with prior evidence that neurodegenerative disease is associated with reduced neural complexity and a diminished repertoire of neural dynamics \citep{jeong2004eeg, stam2005nonlinear, babiloni2016brain}.

Taken together, these findings suggest that healthy aging and neurodegeneration occupy different positions along a common geometric axis. Healthy aging appears to shift neural activity toward a more expansive and less compact representational regime without evidence of collapse. Neurodegenerative disease, in contrast, is associated with collapse and rigidification of representational geometry, accompanied by reduced representational flexibility and a narrower range of viable neural states. The association between intrinsic dimensionality and distributional stability remains robust in healthy individuals and is still detectable in clinical populations, although its spatial extent is reduced in Alzheimer's disease. This pattern suggests that the same geometric principle continues to constrain neural activity under pathology, but in a progressively compressed form.

Several limitations should be acknowledged. First, all analyses were based on EEG, which has limited spatial resolution and reflects mixtures of distributed neural sources. Although this does not undermine the large-scale geometric relationships identified here, future work using source-resolved imaging, invasive electrophysiology, or multimodal recordings will be necessary to localize the circuit-level substrates of representational expansion and collapse. Second, intrinsic dimensionality was estimated using statistical proxies that approximate, but do not exhaustively recover, the true geometry of neural population activity. Developing estimators that are better adapted to noisy, finite, and temporally structured neural data remains an important methodological goal. Third, the present study is correlational. While intrinsic dimensionality shows robust associations with distributional stability across multiple conditions, causal perturbation studies will be required to determine whether and how neural circuits regulate representational dimensionality.

In conclusion, our findings identify intrinsic dimensionality as a unifying geometric principle for understanding the temporal stability of neural representations. By shifting the focus from individual signals to evolving empirical distributions, this work provides a compact framework for understanding how neural population activity remains stable without becoming fixed, and how this balance is systematically altered across cognition, healthy aging, and neurodegenerative disease. More broadly, the present results suggest that changes in representational geometry may provide a principled way to distinguish representational expansion from pathological representational collapse.

\section*{Data availability}

The data cannot be made publicly available upon publication because they contain sensitive personal information. The data that support the findings of this study are available upon reasonable request from the authors.

\section*{Acknowledgements}

This work was supported by the National Natural Science Foundation of China (82371471, 62472206), start-up funds from Shenzhen Bay Laboratory (21290021), Shenzhen Science and Technology Innovation Committee (2022410129, KJZD20230923115221044), GuangDong Basic and Applied Basic Research Foundation (2025A1515011645 to ZC.L.), Shenzhen Doctoral Startup Project (RCBS20231211090748082 to XK.S.), Guangdong Provincial Key Laboratory of Advanced Biomaterials (2022B1212010003), Shenzhen Natural
Science Foundation (JCYJ20240813151223031), and the open research fund of the Guangdong Provincial Key Laboratory of Mathematical and Neural Dynamical Systems, the Center for Computational Science and Engineering at Southern University of Science and Technology.

\section*{Conflict of Interest}

The authors declare that they have no competing interests.

\section*{Ethics approval and consent to participate}

This study was conducted in accordance with the Declaration of Helsinki and was reviewed and approved by the Institutional Review Board of Shenzhen People's Hospital (Shenzhen, China) with the approval number: KY-LL-2020483-02. This study has been approved by the Chinese Clinical Trial Registry (chictr.org.cn identifier ChiCTR2300074261: 2023.08.02, ChiCTR2400094124: 2024.12.17). Informed consent was obtained from the patients for the publication of all images, clinical data and other data included in the manuscript. Data were collected from two institutions—Shenzhen People's Hospital and Tianjin Huanhu Hospital—from December 2023 to December 2024.

\appendix

\bibliographystyle{elsarticle-harv} 
\bibliography{reference.bib}



\section{Mathematical Foundations of Distributional Stability}
\label{app:theory}

\subsection{Neural Representations as Probability Measures}

Let $(X,d)$ be a complete separable metric space representing the neural representational space associated with a given brain region.
For a fixed cognitive condition (task or resting state), neural population activity evolves over time as a sequence of observations
$\{x_t\}_{t=1}^T \subset X$.

We model these observations as samples drawn from an underlying probability measure $\mu$ on $(X,d)$.
Importantly, this formulation does not assume temporal independence of the raw neural signals.
Rather, it treats empirical neural representations as repeated samples from a stable population-level distribution that characterizes the representational geometry associated with the cognitive condition.

This distributional formulation allows representational stability to be defined in terms of distances between probability measures, rather than temporal correlations between individual signals.

\subsection{Measure-Based Intrinsic Dimensionality}

To quantify the geometric complexity of a neural representational distribution, we adopt a measure-based notion of intrinsic dimensionality grounded in covering numbers.
This approach characterizes the effective dimension of regions in which most of the probability mass concentrates, while allowing a small fraction of atypical samples to be ignored.

\begin{definition}[Covering number]
For a subset $S \subset X$ and $\varepsilon > 0$, the $\varepsilon$-covering number of $S$ is defined as
\[
\mathcal{N}_\varepsilon(S) := \min \left\{ N : S \subseteq \bigcup_{i=1}^N B(x_i,\varepsilon),\ x_i \in X \right\},
\]
where $B(x,\varepsilon)$ denotes the closed ball of radius $\varepsilon$ centered at $x$.
\end{definition}

\begin{definition}[Measure covering number]
For a probability measure $\mu$ on $X$ and $\tau \in (0,1)$, the $(\varepsilon,\tau)$-covering number is defined as
\[
\mathcal{N}_\varepsilon(\mu,\tau)
:= \inf \left\{ \mathcal{N}_\varepsilon(S) : \mu(S) \ge 1 - \tau \right\}.
\]
\end{definition}

\begin{definition}[Measure-based intrinsic dimension]
The intrinsic dimension of $\mu$ at scale $\varepsilon$ and tolerance $\tau$ is defined as
\[
d_\varepsilon(\mu,\tau)
:= \frac{\log \mathcal{N}_\varepsilon(\mu,\tau)}{-\log \varepsilon}.
\]
\end{definition}

This definition quantifies how the number of local degrees of freedom required to cover the dominant portion of the distribution scales with resolution.
Low intrinsic dimensionality indicates concentration near a low-dimensional manifold, whereas higher dimensionality reflects more complex geometric structure.

\subsection{Wasserstein Distance and Empirical Measures}

To compare neural representational distributions, we use the $p$-Wasserstein distance.
For $p \ge 1$, the Wasserstein distance between two probability measures $\mu$ and $\nu$ on $(X,d)$ is defined as
\[
W_p(\mu,\nu)
:= \left( \inf_{\pi \in \Pi(\mu,\nu)}
\int_{X \times X} d(x,y)^p \, \mathrm{d}\pi(x,y)
\right)^{1/p},
\]
where $\Pi(\mu,\nu)$ denotes the set of all couplings of $\mu$ and $\nu$.

Given $n$ samples $\{x_i\}_{i=1}^n$ drawn from $\mu$, the empirical measure is defined as
\[
\hat{\mu}_n := \frac{1}{n} \sum_{i=1}^n \delta_{x_i},
\]
where $\delta_x$ denotes the Dirac measure at $x$.

In the present context, Wasserstein distance quantifies the similarity between empirical neural representations obtained from different time segments under the same cognitive condition.

\subsection{Wasserstein Dimension and Convergence Rates}

The rate at which empirical measures converge to the underlying distribution in Wasserstein distance depends on a measure-dependent notion of effective dimensionality.
Following \citet{weed2019sharp}, we define the upper Wasserstein dimension.

\begin{definition}[Upper Wasserstein dimension]
For $p \ge 1$, the upper Wasserstein dimension of a probability measure $\mu$ is defined as
\[
d_p^*(\mu)
:= \inf \left\{ s > 2p :
\limsup_{\varepsilon \to 0}
d_\varepsilon\!\left(\mu, \varepsilon^{sp/(s-2p)}\right)
\le s
\right\}.
\]
\end{definition}

The upper Wasserstein dimension captures the effective dimensionality governing the convergence behavior of empirical measures, while discounting small regions of high geometric complexity that carry negligible probability mass.

\subsection{Empirical Convergence in Wasserstein Distance}

The following result establishes a quantitative link between intrinsic dimensionality and sampling stability.

\begin{theorem}[Empirical Wasserstein convergence {\citep{weed2019sharp}}]
\label{thm:wasserstein_convergence}
Let $\mu$ be a probability measure on $(X,d)$ with finite $p$-th moment, and let $\hat{\mu}_n$ be the empirical measure constructed from $n$ independent samples drawn from $\mu$.
For any $\varepsilon > 0$, letting $s = d_p^*(\mu) + \varepsilon$, there exists a constant $C_{\varepsilon,p} > 0$ such that
\[
\mathbb{E}\big[ W_p(\mu, \hat{\mu}_n) \big]
\le C_{\varepsilon,p} \, n^{-1/s}.
\]
\end{theorem}

This result implies that the expected Wasserstein distance between empirical distributions sampled from the same underlying neural representational distribution is controlled by its intrinsic dimensionality.
Lower-dimensional distributions yield smaller expected representational displacement between samples, whereas higher-dimensional distributions admit greater sampling variability.

\subsection{Interpretation for Neural Representational Stability}

Theorem~\ref{thm:wasserstein_convergence} provides a formal justification for interpreting distributional stability as a geometric property of neural representations.
When neural activity under a fixed cognitive condition is modeled as sampling from a distribution with low intrinsic dimensionality, empirical representations obtained at different times are expected to remain close in Wasserstein distance.
Conversely, higher intrinsic dimensionality implies greater expected variability between empirical samples, even in the absence of systematic drift.

This theoretical result underpins our empirical analyses.
By estimating intrinsic dimensionality from neural data, we obtain a concise geometric descriptor that predicts the stability of neural representational distributions over time.

\subsection{Practical Estimation and Scope}

In practice, intrinsic dimensionality is estimated using local scaling methods such as correlation dimension and maximum likelihood estimators \citep{levina2004maximum}.
These estimators provide empirical proxies for the measure-based intrinsic dimension defined above.

While the convergence result in Theorem~\ref{thm:wasserstein_convergence} is stated for independent samples, our analysis does not require strict temporal independence of neural signals.
Instead, it leverages the empirical observation that neural representations within a fixed cognitive condition exhibit constrained distributional evolution over time.
Under this condition, intrinsic dimensionality remains a meaningful predictor of representational stability, even when samples exhibit weak temporal dependence.

\section{Supplementary Analysis of Constrained Distributional Evolution}
\label{appendix:constrained_channel}

In the main text, we demonstrated that neural representations exhibit 
constrained distributional evolution over time within fixed cognitive states. The primary 
analysis focused on representative frontal channels. To verify that this hypothesis 
is not restricted to specific regions, we here provide additional results for two 
anatomically distinct channels: T7 (temporal) and Oz (occipital).

These channels were selected to probe whether constrained distributional evolution 
persists across cortical regions with different intrinsic dimensionality profiles. 
For each channel, we computed the Wasserstein distance between empirical 
distributions obtained from successive non-overlapping time segments and a fixed 
reference segment, following the same procedure described in Section 3.1.

Across all task conditions, the Wasserstein distance fluctuates over time but does 
not exhibit monotonic growth. Importantly, the displacement remains confined within 
a stable range throughout the recording period. This behavior rules out progressive 
representational drift and confirms that constrained evolution is a global property of neural population activity rather than a region-specific phenomenon.

These supplementary results strengthen the interpretation that neural activity under 
a fixed cognitive condition repeatedly samples from a stable representational 
distribution, with region-dependent but temporally constrained variability.

\begin{figure}[!t]
  \centering
  \includegraphics[width=0.99\linewidth]{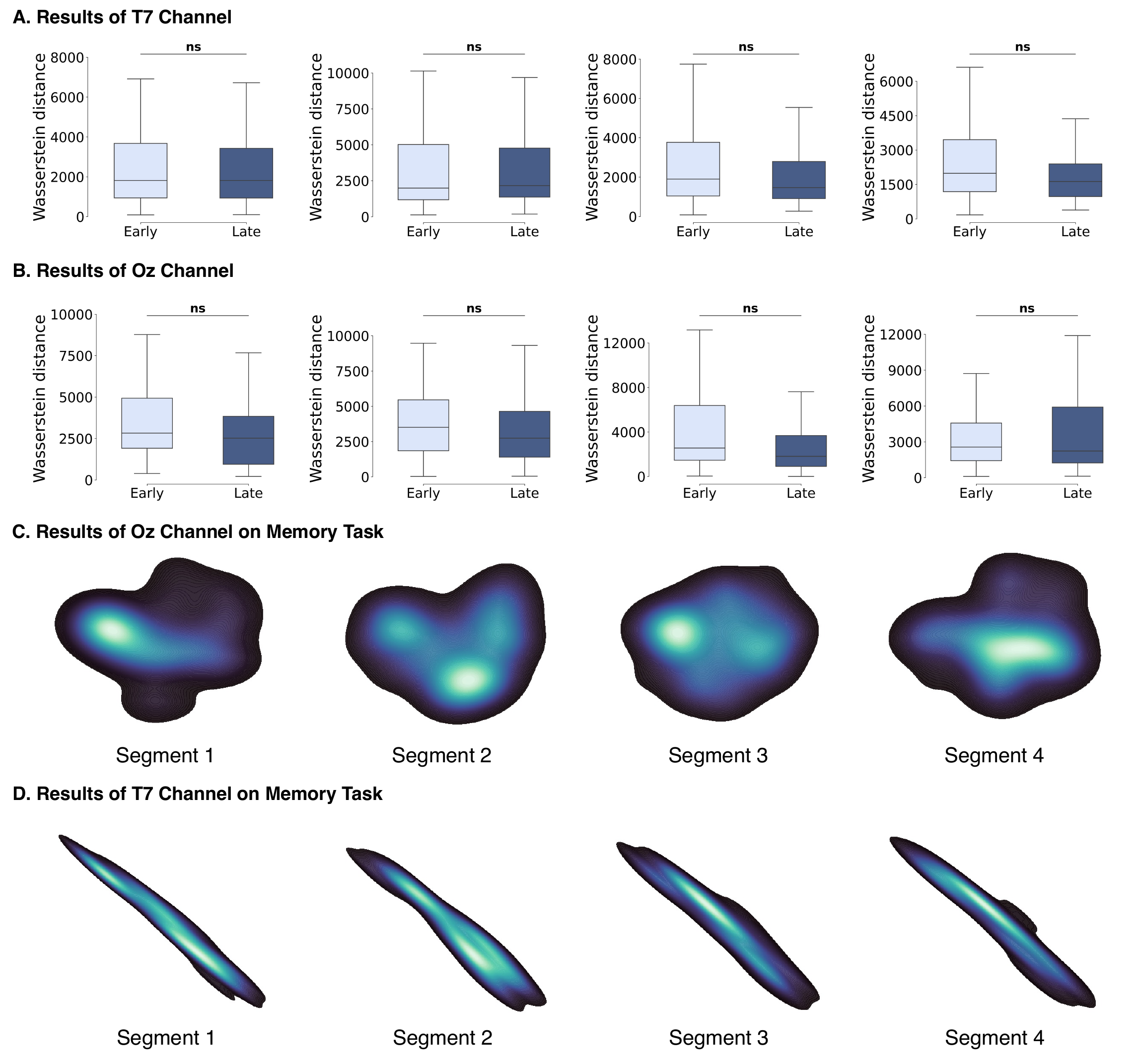}
  \caption{
  \textbf{Constrained distributional evolution in additional cortical channels.}
  Temporal evolution of within-condition Wasserstein distance for the T7 (temporal) 
  and Oz (occipital) channels. For each channel, distances are computed between 
  successive empirical distributions and the initial reference segment. 
  Across cognitive states, Wasserstein distance exhibits non-monotonic but constrained 
  fluctuations, with no evidence of progressive drift. These results demonstrate 
  that constrained distributional evolution is not restricted to frontal regions, 
  but generalizes across cortical areas with distinct intrinsic dimensionality 
  profiles.
  }
  \label{fig:appendix_bound_channel}
\end{figure}

\section{Supplementary Multi-Task Analyses Across Different K Values}
\label{appendix:multitask_validK}

In the main analysis, intrinsic dimensionality was estimated using a fixed number 
of nearest neighbors $K$. To ensure that our conclusions are not driven by a 
particular parameter choice, we repeated the full multi-task analysis across 
multiple values of $K \in \{10, 100\}$.

\begin{figure}[!t]
  \centering
  \includegraphics[width=0.75\linewidth]{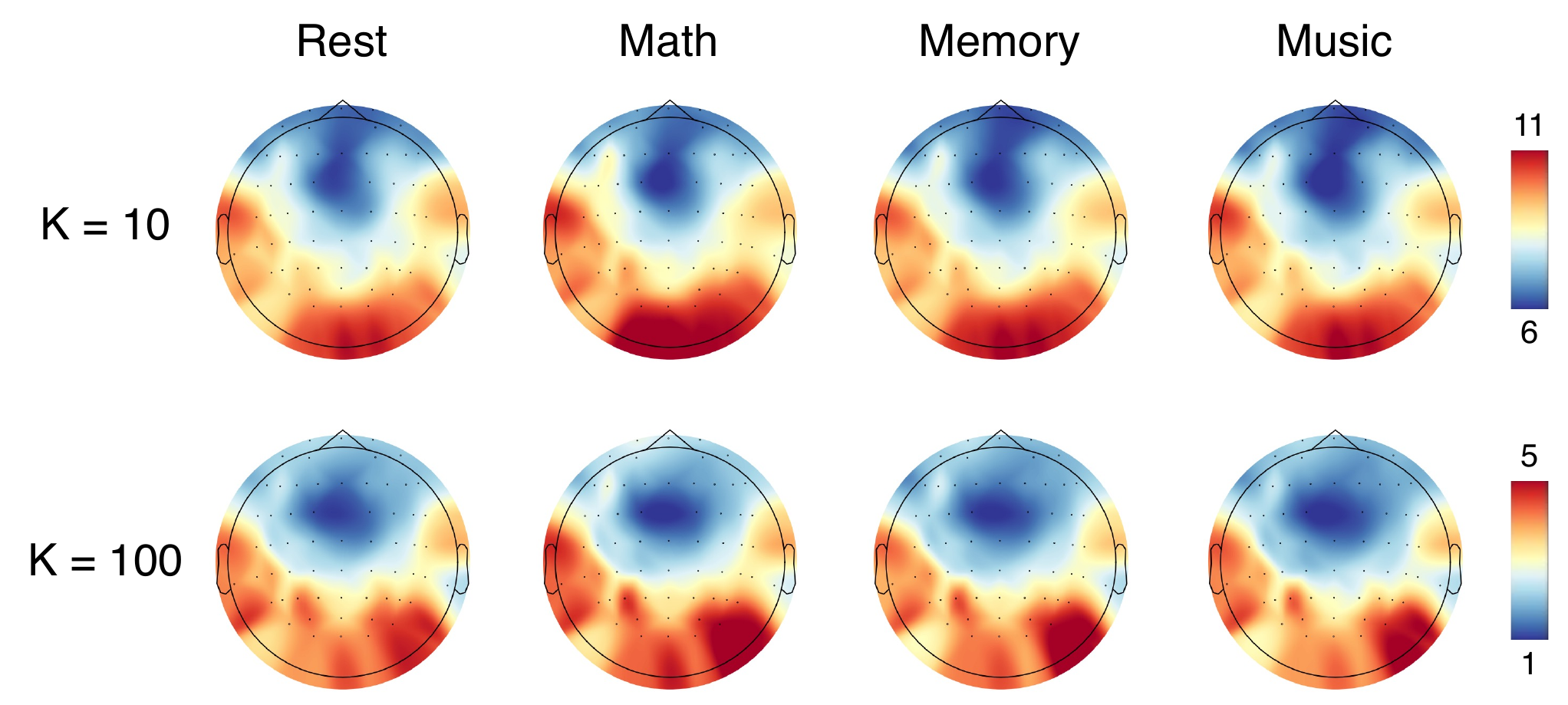}
  \caption{
  \textbf{Whole-brain intrinsic dimensionality across tasks for different $K$.}
  Spatial distributions of intrinsic dimensionality estimated using 
  $K = 10$ and $100$ nearest neighbors. 
  The posterior–anterior gradient is preserved across all parameter settings 
  and cognitive conditions, demonstrating the robustness of the dimensional 
  organization to estimator scale.
  }
  \label{fig:appendix_multitask_dist_validK}
\end{figure}

Figure~\ref{fig:appendix_multitask_dist_validK} shows the spatial distribution of 
intrinsic dimensionality across resting state and cognitive tasks for each $K$. 
Across all parameter choices, the relative ordering of cortical regions remains 
consistent. In particular, posterior and temporal regions operate in higher-dimensional regimes, whereas frontal regions remain comparatively low-dimensional.

This stability across $K$ confirms that the observed spatial gradient of intrinsic 
dimensionality reflects an intrinsic geometric property of neural representations, 
rather than an artifact of estimator hyperparameters.

\section{Supplementary Lifespan Analyses Across Different K Values}
\label{appendix:aging_validK}

To further validate the robustness of the lifespan findings, we repeated the 
aging analysis using multiple values of $K$.

\subsection{Age-Related Changes in Intrinsic Dimensionality}

\begin{figure}[!t]
  \centering
  \includegraphics[width=0.75\linewidth]{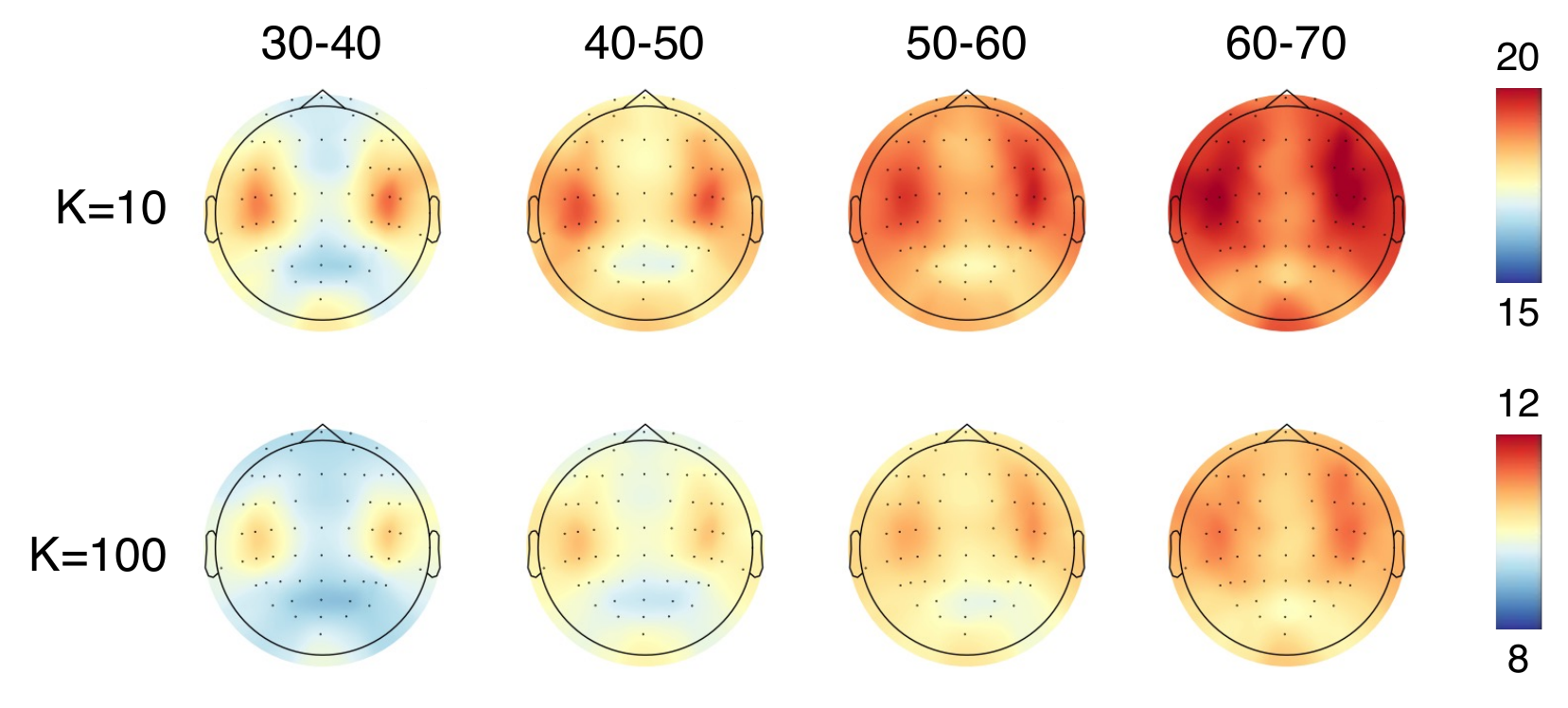}
  \caption{
  \textbf{Age-related increase in intrinsic dimensionality across different $K$.}
  Whole-brain intrinsic dimensionality distributions for 
  $K = 10$ and $100$. 
  Across parameter choices, dimensionality exhibits a consistent 
  upward trend with age, indicating robust representational expansion 
  throughout the adult lifespan.
  }
  \label{fig:appendix_aging_dist_validK}
\end{figure}

Figure~\ref{fig:appendix_aging_dist_validK} shows whole-brain intrinsic 
dimensionality distributions across age groups for different $K$. 
Across all parameter settings, intrinsic dimensionality increases 
systematically with age, and the spatial pattern of expansion remains 
consistent.

This confirms that the observed representational expansion during 
healthy aging is not driven by a specific estimator scale.

\subsection{Robustness of Dimension–Stability Coupling in Aging}

We also examined whether the correlation between intrinsic dimensionality 
and Wasserstein distance remains significant across $K$ in the lifespan dataset.

\begin{figure}[!t]
  \centering
  \includegraphics[width=0.5\linewidth]{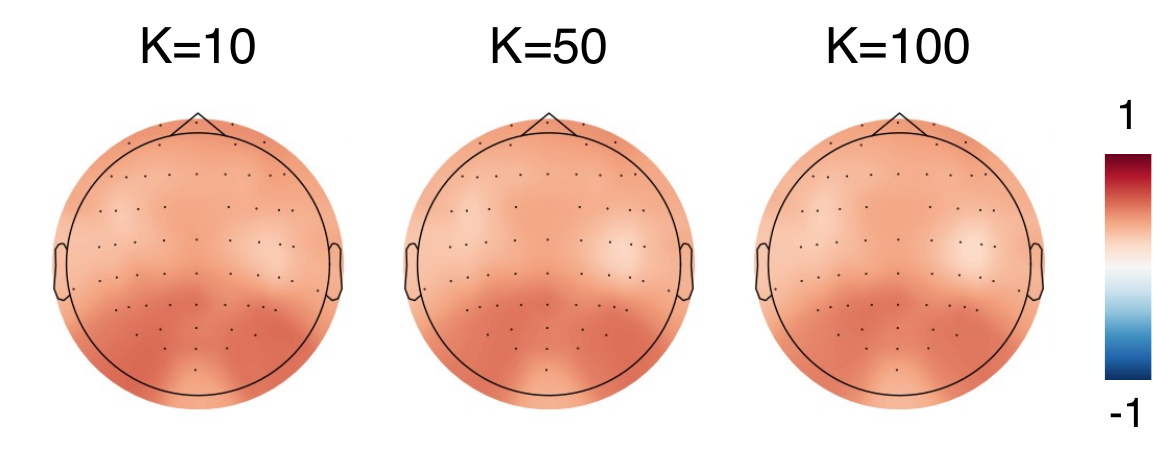}
  \caption{
  \textbf{Robustness of the dimension–stability correlation in aging across $K$.}
  Channel-wise correlations between intrinsic dimensionality and 
  within-condition Wasserstein distance for different values of $K$. 
  Significant positive coupling persists across scales, demonstrating 
  that the geometric relationship remains intact throughout the lifespan.
  }
  \label{fig:appendix_aging_corr_validK}
\end{figure}

As shown in Figure~\ref{fig:appendix_aging_corr_validK}, positive correlations 
are consistently observed across scales. Although effect sizes vary modestly, 
the overall coupling between dimensionality and distributional displacement 
remains statistically significant.

These findings confirm that healthy aging corresponds to a joint expansion 
of intrinsic dimensionality and representational displacement, and that 
their geometric coupling is stable across estimator scales.

\section{Supplementary Clinical Analyses Across Different K Values}
\label{appendix:disease_validK}

Finally, we examined whether the clinical findings remain robust across 
different choices of $K$.

\begin{figure}[!t]
  \centering
  \includegraphics[width=0.9\linewidth]{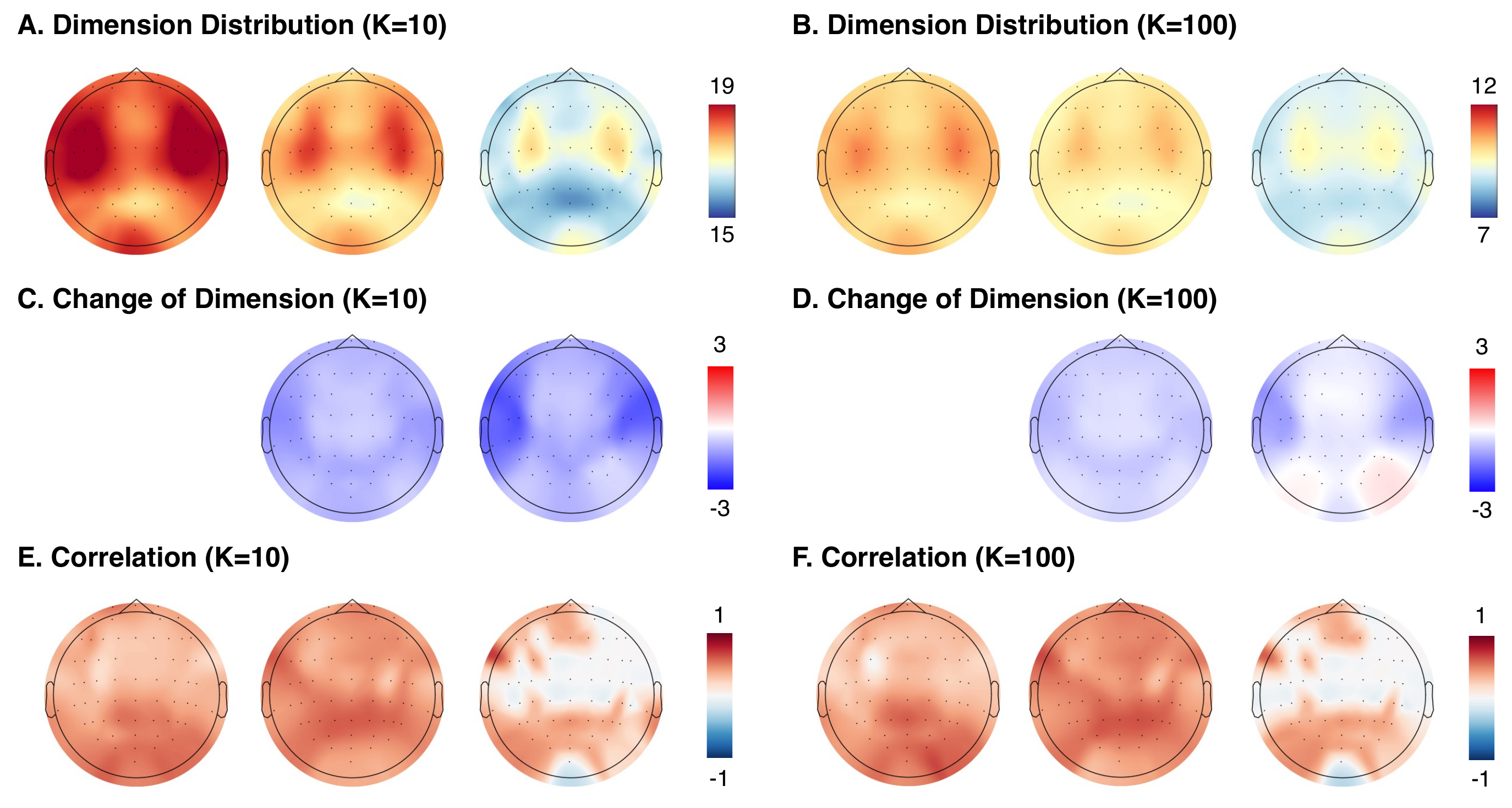}
  \caption{
  \textbf{Intrinsic dimensionality in HC, MCI, and AD across different $K$.}
  Whole-brain intrinsic dimensionality distributions estimated using 
  multiple neighborhood sizes. Across all values of $K$, 
  both MCI and AD groups exhibit reduced dimensionality relative 
  to healthy controls. The positive coupling between intrinsic 
  dimensionality and Wasserstein distance remains present within 
  each group, indicating preservation of the underlying geometric 
  constraint despite pathological shifts.
  }
  \label{fig:appendix_disease_validK}
\end{figure}

Figure~\ref{fig:appendix_disease_validK} shows intrinsic dimensionality 
distributions for healthy controls (HC), mild cognitive impairment (MCI), 
and Alzheimer's disease (AD) across multiple $K$ values.

Across all parameter settings, both MCI and AD groups exhibit 
systematically lower intrinsic dimensionality compared to healthy controls. 
This dimensional collapse is spatially widespread and consistent 
across estimator scales.

In addition, within each group, intrinsic dimensionality remains 
positively correlated with within-condition Wasserstein distance, 
indicating that the geometric constraint linking dimensionality 
and distributional stability persists even under pathological 
conditions.

These results confirm that the representational collapse observed 
in neurodegenerative disease is a robust geometric phenomenon, 
rather than a scale-dependent artifact.

\end{document}